\documentclass[10pt,conference,a4paper,twocolumn]{IEEEtran}
\usepackage{times}
\usepackage{amsmath}
\usepackage{graphicx}
\usepackage{multirow}
\usepackage[none]{hyphenat}
\usepackage[flushleft]{threeparttable}
\usepackage{float}
\usepackage{subfig}
\usepackage{xcolor}
\definecolor{deepgreen}{RGB}{34,139,34} 
\DeclareMathAlphabet{\mathbfit}{OML}{cmm}{b}{it}
\usepackage{cite}
\usepackage{todonotes}
\usepackage{soul}
\usepackage{multicol}
\usepackage{mathrsfs,amssymb}
\usepackage{dblfloatfix}
\usepackage{makecell}
\usepackage{physics}
\usepackage{siunitx}
\usepackage{cite}
\usepackage{amsmath,amssymb,amsfonts}
\usepackage{algorithmic}
\usepackage{graphicx}
\usepackage{textcomp}
\usepackage{xcolor}
\usepackage{algorithm}
\usepackage{comment}
\usepackage{amsmath,amssymb}
\usepackage{url}
\usepackage{hyperref}

\allowdisplaybreaks
\makeatletter

\def\@maketitle{\newpage
\bgroup\par\addvspace{0.5\baselineskip}\centering%
\ifCLASSOPTIONtechnote
   {\bfseries\large\@IEEEcompsoconly{\sffamily}\@title\par}\vskip 1.3em{\lineskip .5em\@IEEEcompsoconly{\sffamily}\@author
   \@IEEEspecialpapernotice\par{\@IEEEcompsoconly{\vskip 1.5em\relax
   \@IEEEtitleabstractindextextbox{\@IEEEtitleabstractindextext}\par
   \hfill\@IEEEcompsocdiamondline\hfill\hbox{}\par}}}\relax
\else
   \vskip0.2em{\EuMWtitlesize\ifCLASSOPTIONtransmag\bfseries\LARGE\fi\@IEEEcompsoconly{\sffamily}\@IEEEcompsocconfonly{\normalfont\normalsize\vskip 2\@IEEEnormalsizeunitybaselineskip
   \bfseries\Large}\@title\par}\vskip1.0em\par
   \ifCLASSOPTIONconference%
      {\@IEEEspecialpapernotice\mbox{}\vskip\@IEEEauthorblockconfadjspace%
       \mbox{}\hfill\begin{@IEEEauthorhalign}\@author\end{@IEEEauthorhalign}\hfill\mbox{}\par}\relax
   \else
      \ifCLASSOPTIONpeerreviewca
         {\@IEEEcompsoconly{\sffamily}\@IEEEspecialpapernotice\mbox{}\vskip\@IEEEauthorblockconfadjspace%
          \mbox{}\hfill\begin{@IEEEauthorhalign}\@author\end{@IEEEauthorhalign}\hfill\mbox{}\par
          {\@IEEEcompsoconly{\vskip 1.5em\relax
           \@IEEEtitleabstractindextextbox{\@IEEEtitleabstractindextext}\par\hfill
           \@IEEEcompsocdiamondline\hfill\hbox{}\par}}}\relax
      \else
         \ifCLASSOPTIONtransmag
           {\@IEEEspecialpapernotice\mbox{}\vskip\@IEEEauthorblockconfadjspace%
            \mbox{}\hfill\begin{@IEEEauthorhalign}\@author\end{@IEEEauthorhalign}\hfill\mbox{}\par
           {\vspace{0.5\baselineskip}\relax\@IEEEtitleabstractindextextbox{\@IEEEtitleabstractindextext}\vspace{-1\baselineskip}\par}}\relax
         \else
           {\lineskip.5em\@IEEEcompsoconly{\sffamily}\sublargesize\@author\@IEEEspecialpapernotice\par
           {\@IEEEcompsoconly{\vskip 1.5em\relax
            \@IEEEtitleabstractindextextbox{\@IEEEtitleabstractindextext}\par\hfill
            \@IEEEcompsocdiamondline\hfill\hbox{}\par}}}\relax
         \fi
      \fi
   \fi
\fi\par\addvspace{0.0\baselineskip}\egroup}

\def\EuMWtitlesize{\@setfontsize{\EuMWtitlesize}{24}{24pt}}
\def\EuMWauthorsize{\@setfontsize{\EuMWauthorsize}{11}{11pt}}
\def\EuMWaffilsize{\@setfontsize{\EuMWaffilsize}{10}{10pt}}
\def\EuMWcaptionsize{\@setfontsize{\EuMWcaptionsize}{9}{10pt}}
\def\EuMWbibsize{\@setfontsize{\EuMWbibsize}{8}{10pt}}

\def\@IEEEauthorblockNstyle{\EuMWauthorsize\@IEEEcompsocnotconfonly{\sffamily}\@IEEEcompsocconfonly{\large}}
\def\@IEEEauthorblockAstyle{\EuMWaffilsize\@IEEEcompsocnotconfonly{\sffamily}\@IEEEcompsocconfonly{\itshape}\@IEEEcompsocconfonly{\large}}
\def\@IEEEauthordefaulttextstyle{\EuMWauthorsize\@IEEEcompsocnotconfonly{\sffamily}\sublargesize}

\def\thebibliography#1{\section*{\refname}%
    \addcontentsline{toc}{section}{\refname}%
    \EuMWbibsize\@IEEEcompsocconfonly{\small}\vskip 0.3\baselineskip plus 0.1\baselineskip minus 0.1\baselineskip
    \list{\@biblabel{\@arabic\c@enumiv}}%
    {\settowidth\labelwidth{\@biblabel{#1}}%
    \leftmargin\labelwidth
    \advance\leftmargin\labelsep\relax
    \itemsep \IEEEbibitemsep\relax
    \usecounter{enumiv}%
    \let\p@enumiv\@empty
    \renewcommand\theenumiv{\@arabic\c@enumiv}}%
    \let\@IEEElatexbibitem\bibitem%
    \def\bibitem{\@IEEEbibitemprefix\@IEEElatexbibitem}%
\def\newblock{\hskip .11em plus .33em minus .07em}%
\ifCLASSOPTIONtechnote\sloppy\clubpenalty4000\widowpenalty4000\interlinepenalty100%
\else\sloppy\clubpenalty4000\widowpenalty4000\interlinepenalty500\fi%
    \sfcode`\.=1000\relax}

\long\def\@makecaption#1#2{%
\ifx\@captype\@IEEEtablestring%
\par\@IEEEtabletopskipstrut
\else
\@IEEEfigurecaptionsepspace
\fi
\setbox\@tempboxa\hbox{\normalfont\footnotesize {#1.}\nobreakspace\nobreakspace #2}%
\ifdim \wd\@tempboxa >\hsize%
\setbox\@tempboxa\hbox{\normalfont\footnotesize {#1.}\nobreakspace\nobreakspace}%
\parbox[t]{\hsize}{\normalfont\footnotesize\noindent\unhbox\@tempboxa#2}%
\else
\ifCLASSOPTIONconference \hbox to\hsize{\normalfont\footnotesize\hfil\box\@tempboxa\hfil}%
\else \hbox to\hsize{\normalfont\footnotesize\box\@tempboxa\hfil}%
\fi\fi
\ifx\@captype\@IEEEtablestring%
\@IEEEtablecaptionsepspace
\else
\fi}

\newlength\tablecaptiontotableskip
\newlength\figuretocaptionskip
\def\@IEEEfigurecaptionsepspace{\vskip\figuretocaptionskip\relax}%
\def\@IEEEtablecaptionsepspace{\vskip\tablecaptiontotableskip\relax}%

\def\abstract{\normalfont%
\@IEEEabskeysecsize\bfseries\textit{\abstractname}\,\bfseries\textit{---}\,%
\@IEEEgobbleleadPARNLSP}%

\def\IEEEkeywords{\normalfont%
\@IEEEabskeysecsize\bfseries\textit{\IEEEkeywordsname}\,\bfseries\textit{---}\,%
\@IEEEgobbleleadPARNLSP}%
\def\endIEEEkeywords{\relax\vspace{0.67ex}%
\par\if@twocolumn\else\endquotation\fi%
\normalsize\normalfont}%

\def\@IEEEauthorblockNtopspace{0ex}
\def\@IEEEauthorblockAtopspace{1mm}
\def\IEEEkeywordsname{Keywords}
\def\subsubsection{\@startsection{subsubsection}{3}{\z@}{1.5ex plus 1.5ex minus 0.5ex}%
{0.7ex plus .5ex minus 0ex}{\normalfont\normalsize\itshape}}%
\newlength{\CPheadmatchindent}%
\def\@seccntformat#1{\hbox to\CPheadmatchindent{\csname the#1dis\endcsname}\hskip 0.1em \relax}
\IEEEilabelindentA \parindent
\IEEEilabelindent \IEEEilabelindentA
\IEEEelabelindent \parindent
\IEEEdlabelindent \parindent
\IEEElabelindent \parindent
\makeatother

    \def\BibTeX{{\rm B\kern-.05em{\sc i\kern-.025em b}\kern-.08em T\kern-.1667em\lower.7ex\hbox{E}\kern-.125emX}}
\begin{document}
\pagestyle{plain}

	\addtolength{\oddsidemargin}{0.08in}
	\addtolength{\evensidemargin}{0.08in}

	\addtolength{\topmargin}{0.08in}
	\raggedbottom
	\title{Parasitic Circus: On the Feasibility of Golden-free PCB Verification}
	\author{
		Maryam Saadat Safa, Patrick Schaumont and Shahin Tajik\\
		Department of Electrical and Computer Engineering,\\ 
        Worcester Polytechnic Institute, Worcester, MA, USA\\
		\{msafa, pschaumont, stajik\}@wpi.edu

}

	\maketitle
	%
	%

\begin{abstract}
Printed circuit boards (PCBs) are an integral part of electronic systems.
Hence, verifying their physical integrity in the presence of supply chain attacks (e.g., tampering and counterfeiting) is of utmost importance. 
Recently, tamper detection techniques grounded in impedance characterization of PCB's Power Delivery Network (PDN) have gained prominence due to their global detection coverage, non-invasive, and low-cost nature. 
Similar to other physical verification methods, these techniques rely on the existence of a physical golden sample for signature comparisons. 
However, having access to a physical golden sample for golden signature extraction is not feasible in many real-world scenarios.
In this work, we assess the feasibility of eliminating a \emph{physical} golden sample and replacing it with a simulated golden signature obtained by the PCB design files.
By performing extensive simulation and measurements on an in-house designed PCB, we demonstrate how the parasitic impedance of the PCB components plays a major role in reaching a successful verification.
Based on the obtained results and using statistical metrics, we show that we can mitigate the discrepancy between collected signatures from simulation and measurements.

    \end{abstract}
    

\maketitle

\section{Introduction}
Printed circuit boards (PCBs) serve as the essential foundation for virtually all electronic systems. 
They house a variety of microelectronic components, starting from basic discrete devices like diodes, transistors, resistors, and capacitors, all the way to sophisticated integrated circuits (ICs) such as microprocessors, field-programmable gate arrays (FPGAs), and memory modules.
The globalized supply chain for PCB manufacturing and assembly leaves PCBs vulnerable to attacks, such as tampering~\cite{big_hack} and counterfeiting~\cite{fake_cisco}.
Therefore, it is critical to verify PCB's physical integrity before their deployment in the field.
Numerous PCB tamper/counterfeit detection methods have been proposed ranging from inspection using imaging techniques (e.g., X-ray~\cite{botero2021hardware} and visual inspection~\cite{chaudhary2017automatic}) to behavioral analysis using side-channels~\cite{piliposyan2022pcb}. 
However, most of these approaches suffer from lack of scalability, high cost, and limited detection coverage.
To mitigate these shortcomings, novel non-invasive techniques based on impedance characterization of PCB's power delivery network have been introduced~\cite{mosavirik2022scatterverif,werner2022detection,zhu2023pdnpulse,safa2023counterfeit,mosavirik2023impedanceverif}.
As any tampering attempt on the PCB or IC will lead to changes in the equivalent impedance of the PDN, the physical monitoring of it provides a holistic solution for determining whether the system’s integrity has been violated.
While these methods are both precise and efficient, similar to other physical verification methods, the dependence on golden samples for comparison poses a significant challenge.
Acquiring these golden samples is notably difficult as a trustworthy PCB assembly factory should exist to manufacture them.
Naturally, such a condition might not be met in real-world scenarios and only the design files of the system (e.g., PCB netlist, bill of material, and IC package specifications) are accessible to the verifier.
Hence, detection methods that do not rely on golden samples are preferred.

\begin{figure}[t!]
     	\centering \noindent
     	\includegraphics[width=0.5\textwidth]{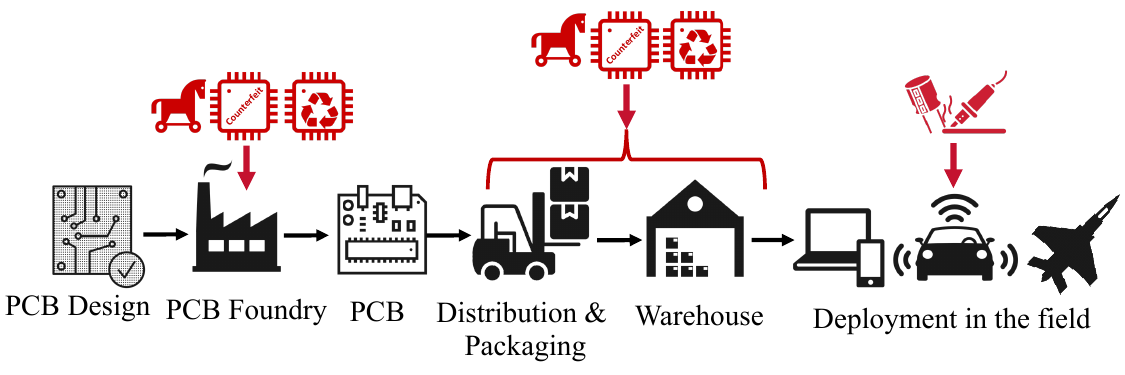}
      \vspace*{-5mm}
     	\caption{An illustration of the PCB's supply chain reveals the possible risks, such as hardware Trojans and counterfeit or recycled parts, that may be inserted at each stage of PCB’s supply chain.}
     	\label{Fig_0}
        \end{figure}



There has been a few attempts in the literature to eliminate the need for the physical golden sample requirement~\cite{bhattacharyay2022vipr,chowdhury2023golden,krishnamurthy2022multi}.
However, the tamper detection capabilities of such framework are confined to the PCB’s circuit or chip’s firmware and, hence, cannot detect physical modifications, e.g., adding an extra via to the PCB or modifying the PCB materials.
A primary challenge in the physical characterization arises from the slight variances in hardware due to the existing manufacturing process variations.
Motivated by the constraints and challenges highlighted previously, we are compelled to confront a pivotal question: \emph{is there a possibility to develop a generic method capable of detecting physical tamper events without relying on a physical golden sample, and if so, under what circumstances?}

\noindent{\bfseries Our Contribution:}
In this work, we demonstrate the feasibility of utilizing PCB design files to generate an estimated golden signature, which is then compared to the measured signature of untrusted boards.
To validate this approach, we simulate the trusted PCB layout using a sophisticated tool and extract an RF signature characterized by the scattering parameters.
We model attacks as additions/removals of components to the PCB's PDN circuit or replacing them with fake parts.
Finally, we compare the trusted simulated data to the measured signature derived from the same physical layout using a similarity measure called dynamic time warping (DTW).
Using the DTW metric, we demonstrate that genuine and tampered or counterfeited PCBs can be reliably detected using the simulated golden signatures. 
Consequently, we provide evidence to support the assertion that \emph{it is feasible to use the simulated signature as the golden signature, provided that the approximate values of the parasitic impedance of the PCB components are known by the verifier.}
We also demonstrate that through the application of this tampering attack detection method, \emph{it becomes possible to identify attacks related to the PDNs that are not directly connected to accessible ports}. 
This detection capability extends beyond conventional means, reaching areas typically considered outside the scope of the testing.


\section{Background}\label{background}

\begin{figure}[t!]
     	\centering \noindent
     	\includegraphics[width=9cm]{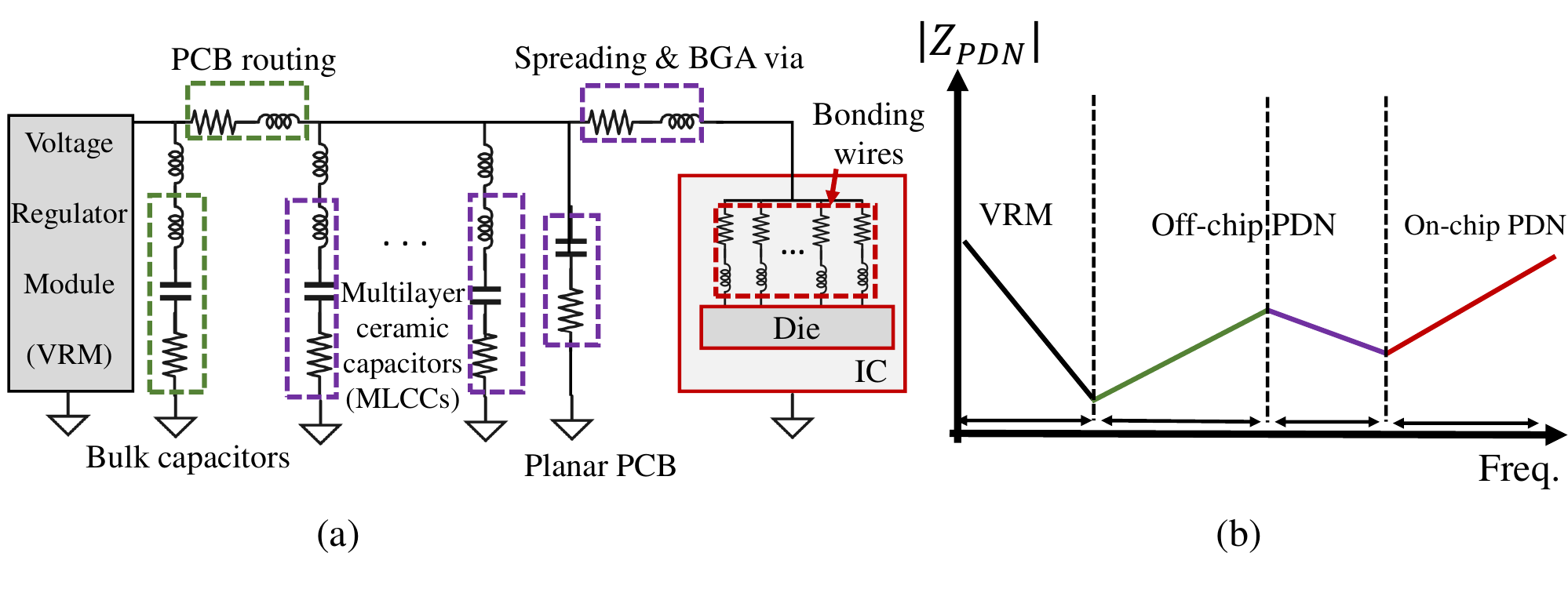}
      \vspace*{-5mm}
     	\caption{(a) The equivalent circuit of the PDN of an electronic board~\cite{mosavirik2023impedanceverif}. (b) The amplitude of the impedance profile of an electronic board over frequency.}
     	\label{Fig_1}
        \end{figure}
\subsection{Power Delivery Network (PDN)}\label{sec:pdn}
The power delivery network (PDN) serves the crucial role of providing a stable and sufficient power supply to various modules on the PCB, see Fig.~\ref{Fig_1}(a). 
In complex PCB designs, different chips and components have diverse power distribution requirements, including specific supply voltage levels, maximum load currents, and voltage noise margins. 
To fulfill these requirements, the PDN employs voltage regulator modules (VRMs) arranged in a tree structure, establishing multiple voltage domains.
The off-chip component typically determines the $Z_{PDN}$ (PDN impedance) up to frequencies in the tens of MHz range. However, at higher frequencies, the impedance is primarily influenced by the on-chip PDN, see Fig.~\ref{Fig_1}(b). 
The primary reason behind such behavior is the existing parasitic impedance of various PCB components, which will be explained in the following subsection.


\subsection{The Impact of Component's Parasitic}\label{sec:parasitic}
\begin{figure}[t!]
     	\centering \noindent
     	\includegraphics[width=6.5cm]{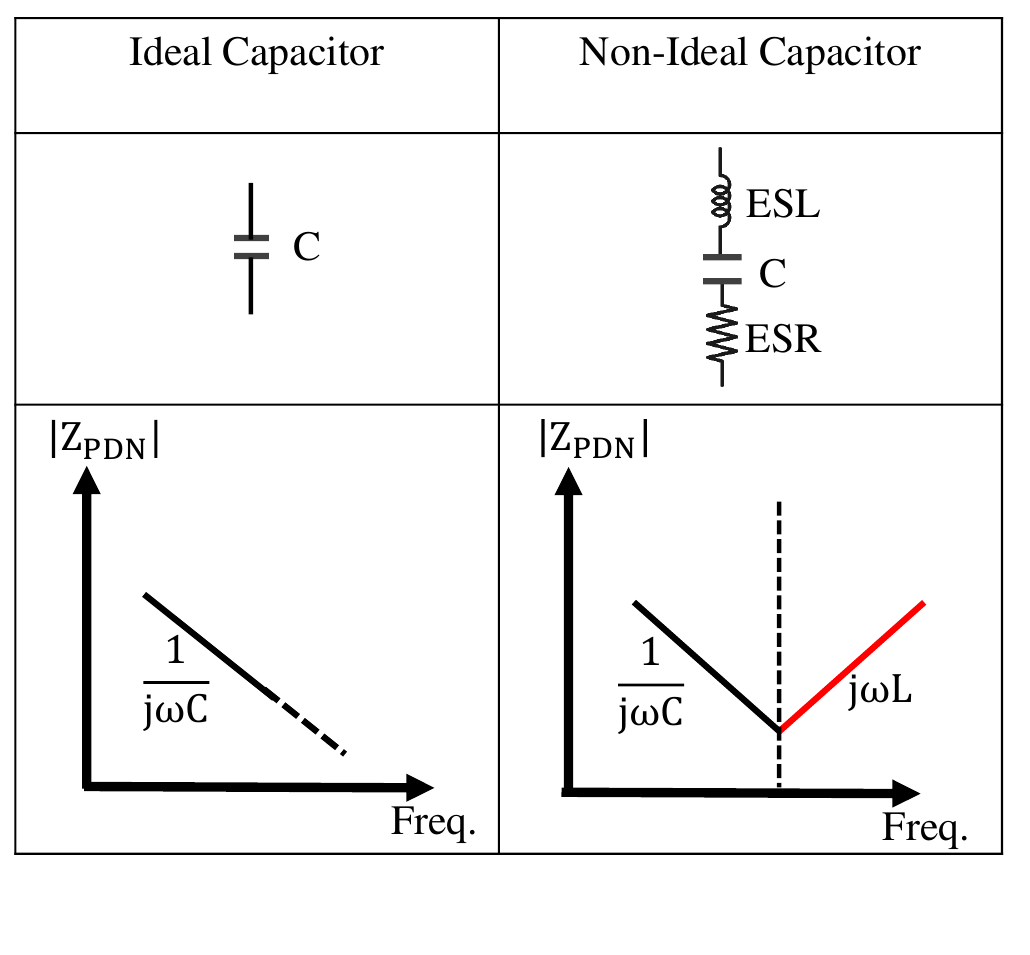}
      \vspace*{-5mm}
      \caption{Impedance profile of an ideal and non-ideal capacitor in log scale.}
     	\label{Fig_4}
        \end{figure}

The real-world PCB components contain parasitics making them behave differently compared to their ideal models.
For instance, in addition to the capacitve behavior, capacitors manifest resistive and inductive characteristics as well, commonly referred to as Equivalent Series Resistance (ESR) and Equivalent Series Inductance (ESL), respectively.
The introduction of components, such as non-ideal capacitors as illustrated in Fig.~\ref{Fig_4}, is modeled by incorporating an RLC branch into the circuit model and introduces a resonance frequency into system.
The resonance frequency of such RLC circuit can be obtained as follows.
\begin{equation}\label{resonance_freq}
f=\dfrac{1}{2\pi\sqrt{LC}}
\end{equation}


Eq.~\ref{resonance_freq} demonstrates that as either capacitance or inductance increases, the resonance frequency decreases.
By adding a component to the PDN, the equivalent capacitance and inductance of the system changes, and consequently, the resonance frequency also shifts in the frequency domain.
Accurate estimation of parasitic impedance is essential in our study, as we are analyzing the impact of tampering (e.g., addition or removal of components). 
If the approximate values of parasitic are not known, the simulated signatures could differ significantly from the measured signatures. 

\subsection{Impedance Characterization using Scattering Parameters}\label{sec:scattering}
\begin{figure}[t!]
     	\centering \noindent
     	\includegraphics[width=7.2cm]{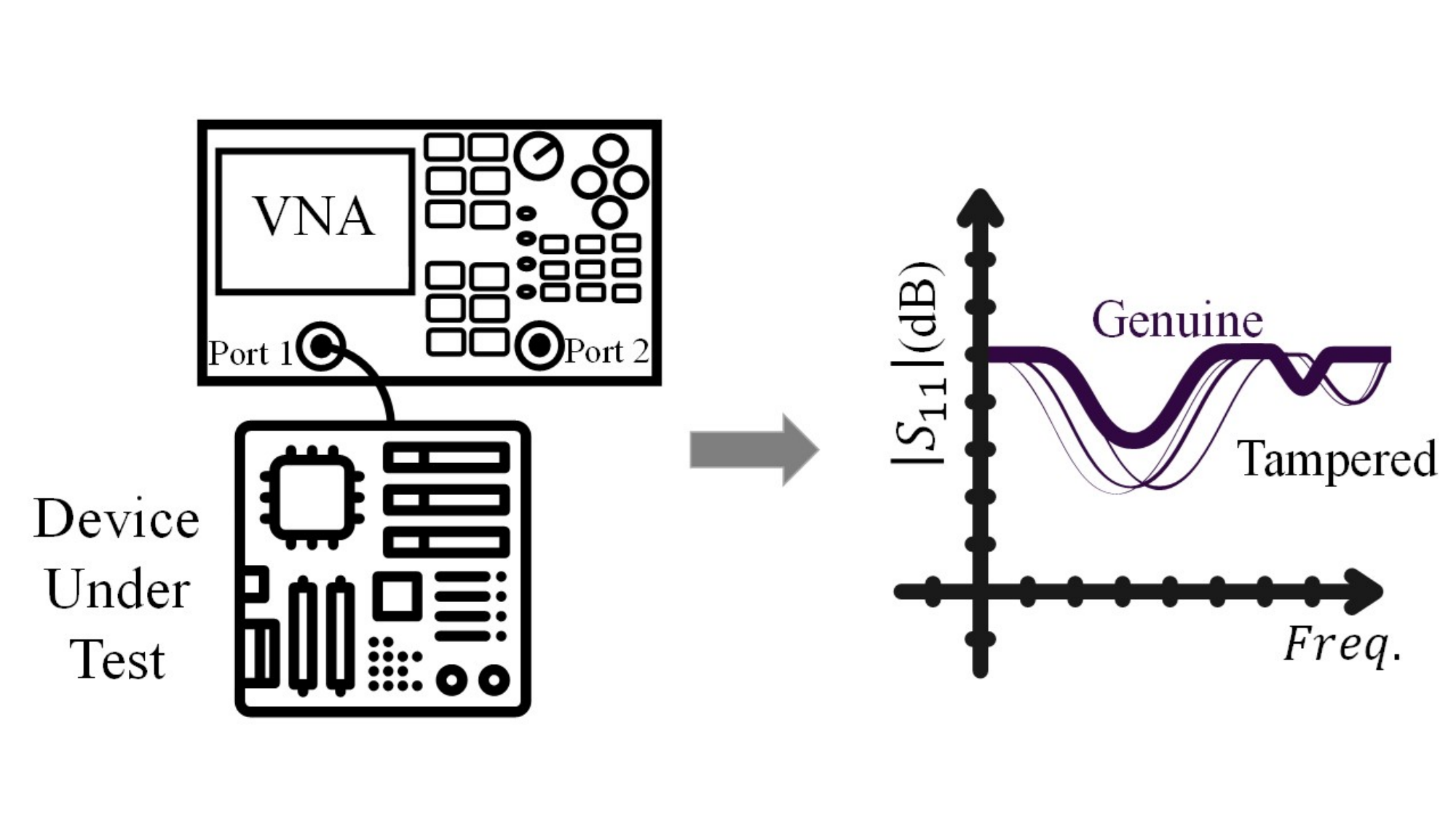}
      \vspace*{-5mm}
     	\caption{Hardware signature extraction based on reflection method.}
     	\label{Fig_2}
        \end{figure}

\begin{figure}[t!]
     	\centering \noindent
     	\includegraphics[width=8cm]{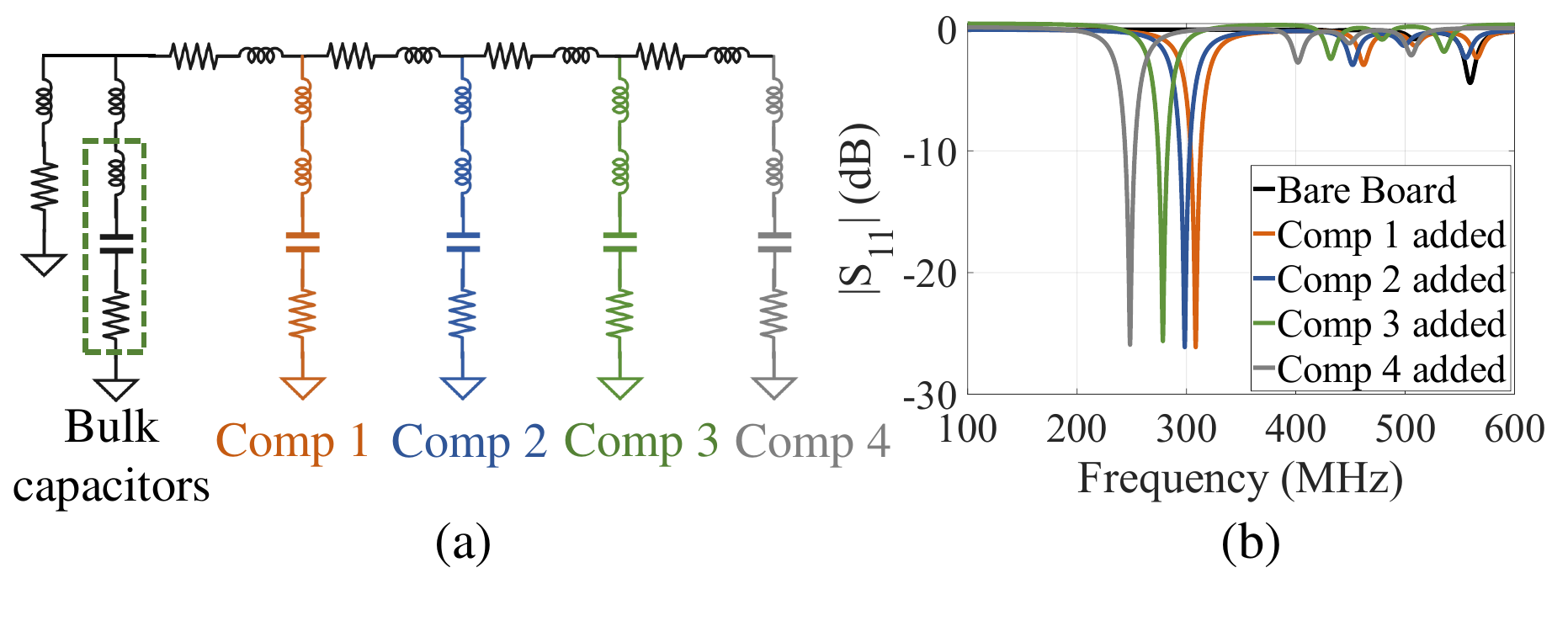}
      \vspace*{-5mm}
     	\caption{(a) PDN circuit model of the PCB. (b) The simulated $|S_{11}|$ displays the impedance contribution of each component. In each step, ranging from 1 to 4, one component is added.}
     	\label{Fig_3}
        \end{figure}

To evaluate the PDN's impedance, we utilize S-parameters (Scattering).
Given the complex nature of an electronic board, it can be modeled as either a single-port or multi-port network. 
The transmitted and reflected power of signals entering and leaving the PDN at various frequencies is measured using a Vector Network Analyzer (VNA).
Using the VNA, we inject sine waves into the PCB for each frequency sample and capture the signal's reflective response from the PDN. 
These sine waves in frequency domain analysis are characterized by their frequency, amplitude, and phase. 
Our study focuses on leveraging the reflected amplitude response ($|S_{11}|$), obtainable directly from the VNA, see Fig.~\ref{Fig_2}.
The relationship between the impedance of the device under test (DUT), represented as $Z_{DUT}$, and the reflection coefficient $S_{11}$, is expressed in Eq.~\ref{Conversion_to_Z}:

\begin{equation}\label{Conversion_to_Z}
Z_{DUT}=Z_0\dfrac{1+S_{11}}{1-S_{11}}
\end{equation}

where $Z_{0}$ is the characteristic impedance of the connecting cables to the VNA.
The relationship expressed in Eq.~\ref{Conversion_to_Z} clarifies that the reflection coefficient provides valuable insights into the impedance characteristics of the system. 
Fig.~\ref{Fig_3} illustrates the impact of tampering with the PCB.
In this case, the addition of capacitors to the PDN causes change in the PDN's impedance, and thus, it affects the $|S_{11}|$.

\section{Methodology}
\subsection{Threat Model}\label{sec:Methodology_general}   
In our threat model, we make the following assumptions.
Firstly, we assume that the attacker can make physical alterations to the PCB at any point in its life cycle, encompassing fabrication, integration, distribution, and repair stages.
The adversary also possesses the capability to perform various modifications at the PCB level. 
These PCB modifications could involve changing the substrate material or drilling into it, as well as adding, removing, modifying, or replacing components of the PCB.
The goal of such tampering could be creating counterfeit or cloned versions of legitimate products, introducing malicious functionality, or embedding backdoors.
Secondly, we assume that the verifier has only access to the design files of the PCB (e.g., PCB netlist, PCB layout, bill of material, and IC package specifications) and the ability to obtain the impedance signature of the PCB's PDN through simulation.
Moreover, the verifier is given a population of genuine and tampered boards for verification, and hence, she can cluster the samples by comparing their signatures to the simulated golden signature and setting a threshold.
The verification does not require control over specific parts of the supply chain.


\subsection{Dynamic Time Warping (DTW)}
Due to existing manufacturing process variation, the parasitic impedance of identical PCB's components varies.
As discussed in the Sect.~\ref{sec:parasitic}, these variations lead to shifts in resonance frequencies and the amplitude of the entire S-parameter profile.
Naturally, such shifts in the signature profiles of identical samples makes the comparison of genuine signatures (generated from simulation and measurement) challenging and could lead to false alarms.
While these shifts exist, the overall pattern of the signatures over frequency remains similar.
To measure such similarities and reduce the impact of such consistent shifts, we deploy Dynamic Time Warping (DTW), which is a similarity measure between time series~\cite{vintsyuk1968speech,sakoe1978dynamic}.



We consider ${\mathcal{S}_{11}}^{Sim}$ and ${\mathcal{S}_{11}}^{Meas}$ as \emph{amplitude} vectors of simulated and measured $S_{11}$ parameters, respectively.
We define the DTW distance for ${\mathcal{S}_{11}}^{Sim}$ and ${\mathcal{S}_{11}}^{Meas}$ as follows,

\begin{multline}\label{DTW}
DTW_{q}(\mathcal{S}_{11}^{Sim},\mathcal{S}_{11}^{Meas}) =\\
\min_{\delta\in A(\mathcal{S}_{11}^{Sim},\mathcal{S}_{11}^{Meas})} (\sum_{i,j\in\delta} d(\mathcal{S}_{11}^{Sim},\mathcal{S}_{11}^{Meas})^q)^{\frac{1}{q}}
\end{multline}

where, an alignment path $\delta$ is a sequence of index pairs and $A(\mathcal{S}_{11}^{Gen},\mathcal{S}_{11}^{Tamp})$ is the set of all admissible paths.

\subsection{Golden-free Tamper Detection Method}
This section discusses the overall framework for golden-free PCB verification.
The principal basis of the proposed golden-free detection methodology is to rely on the simulated traces of $|S_{11}|$ as a golden signature and compare it with the measured traces collected from PCBs under test.
In the first phase, a trusted PCB design file is employed to generate the golden sample signature.
The process involves importing and extracting the electrical characteristics of the PCB from its design file, followed by exporting the S-parameter signature of the PCB using simulation software. 
The PCB's circuit is first segmented into multiple PDNs to reduce its complexity.
Next, the verifier defines ports associated with each PDN, simulates the circuit, and extracts the $|S_{11}|$ parameter. 

\begin{figure}[t!]
     	\centering \noindent
     	\includegraphics[width=6.8cm]{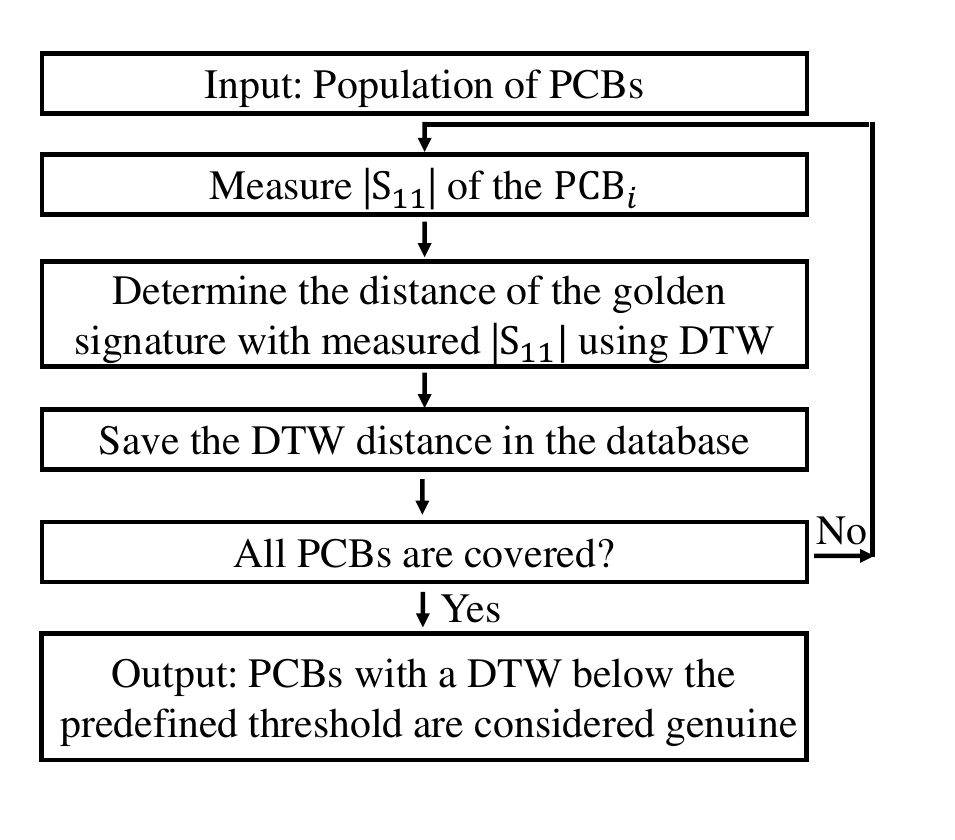}
      \vspace*{-5mm}
     	\caption{The main steps of the proposed verification method.}
     	\label{Fig_6}
        \end{figure}

In the second phase, the verifier performs $|S_{11}|$ measurements using a VNA on the PDNs of a population of PCB samples.
Afterward, the verifier will apply the DTW metric on the generated simulated golden signature and each of collected measured signatures.
If the DTW score is below a predefined threshold, the test will be passed, and the sample is verified as genuine.
Otherwise, the test fails and the sample will be considered dissimilar.
The verifier should set the threshold during the design phase of the PCB.
Depending on the applications in which the PCB will be deployed, the verifier can set various tolerances for components' parasitic.
The flowchart in Fig.~\ref{Fig_6} outlines the major steps that the verifier must undertake during the test phase.
Note that the dissimilarity of the samples does not necessarily point to a tamper event, and finding the root-cause of the signature deviation requires further investigations.



\section{Experimental Setup}
\subsection{Device Under Test (DUT)}\label{sec:dut}
\begin{figure}[t!]
     	\centering \noindent
     	\includegraphics[width=9cm]{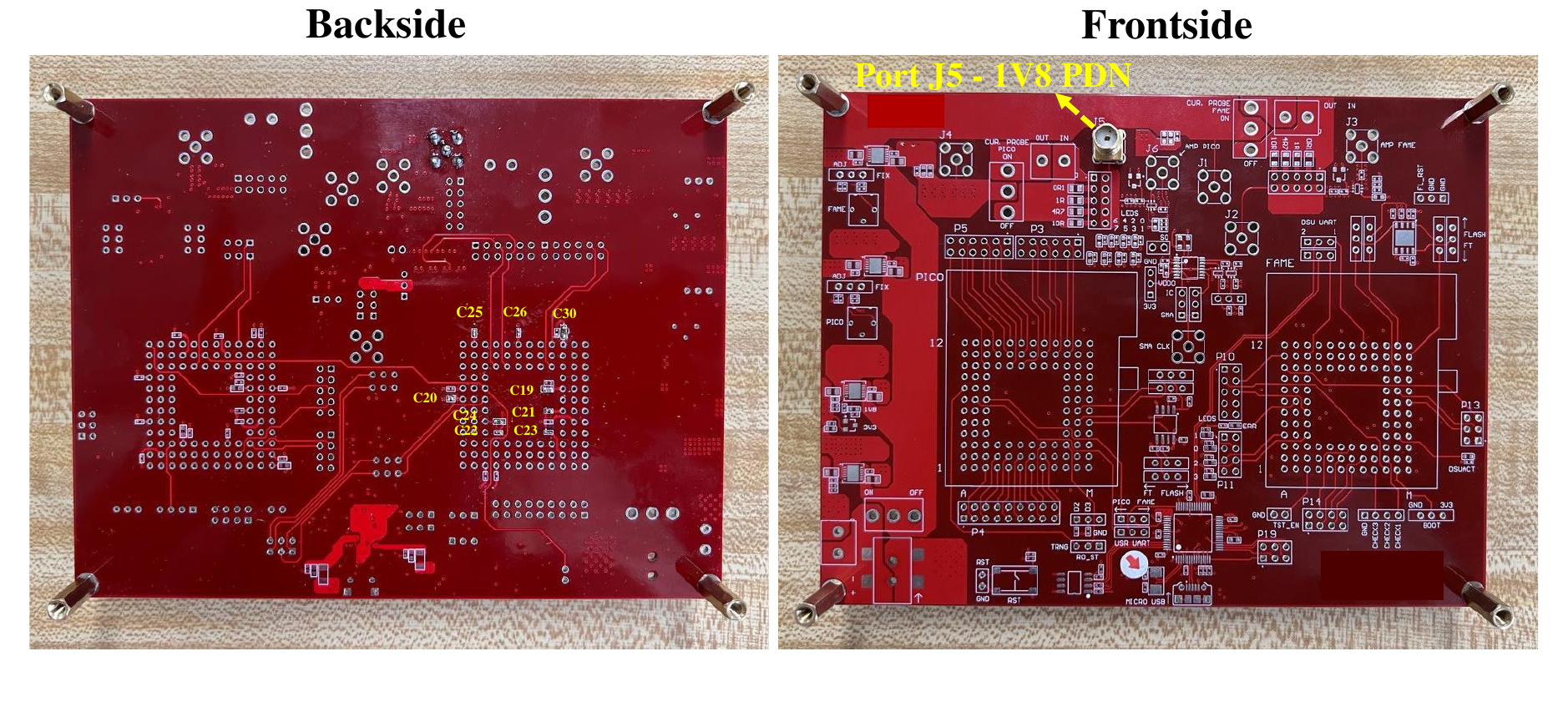}
      \vspace*{-5mm}
     	\caption{Front- and backside of the board under test depicting the circuitry and port J5 of the PDN under test (1V8). Capacitors integrated into the PDN are highlighted.}
     	\label{Fig_7}
        \end{figure}

For our experimental setup, we utilized an in-house designed PCB shown in Fig.~\ref{Fig_7}.
This board contains three distinct and isolated PDNs named 1V8, 3V3 and 5V.
The board, made of FR4 epoxy substrate, consists of 242 components, including capacitors, resistors, ICs, LEDs, SMA ports, headers, traces, and vias.
In this paper, we perform our measurements on the 1V8 PDN and systematically add the components associated with this PDN as shown in Table.~\ref{capacitors}. 
The J5 port, which is connected to the VNA via an SMA connector and gives us direct access to the PDN under test.

\subsection{Simulation Setup}
We used ANSYS SIwave 2023 R2, which is a powerful 2.5D electromagnetic (EM) simulation tool that combines the finite element method (FEM) and the method of moments (MOM)~\cite{ansys_siwave_2023}. 
It utilizes a hybrid solver with a 2-D triangular mesh, enabling it to handle intricate PCB layouts effectively. 
The tool is capable of solving complex layouts, including traces, planes, and through-hole vias, as well as accounting for the effects of metal thickness and dielectric thickness~\cite{siwave}. 
To efficiently utilize computing resources and achieve accurate results, we employ dynamic linking of ANSYS SIwave and HFSS for comprehensive system-level electromagnetic interference (EMI) analysis.

\subsection{Measurement Setup}
We employed the Mini-circuits eVNA-63+ as our Vector Network Analyzer (VNA), which offers a wide operating bandwidth between 300 kHz to 6 GHz. 
To establish a direct connection between the VNA and the DUT, we utilized the CBL-2FT-SMNM+ cable, which is a shielded precision test cables.
These cables feature male SMA connectors on the side of the DUT, thereby facilitating a seamless connection without the requirement for additional adaptors.
For precise calibration, we followed the industry-standard Open-Short-Load (OSL) calibration technique.
Calibration was performed until the SMA connection plane on the board, ensuring accurate and reliable measurements for one-port reflection and impedance analysis.

Our measurements were conducted within a bandwidth of 1 MHz to 1 GHz.
To ensure optimal spectral resolution, we configured the VNA to employ 5000 equally-spaced frequency samples.
To achieve the desired measurement accuracy, we configured the VNA with a 10 kHz Intermediate Frequency (IF) bandwidth and set the output power level to 5 dBm.

\begin{table}
\centering
 \caption{Capacitors utilized in the experiments: all capacitors are connected to the 1V8 PDN, with the exception of $C_{30}$ which is connected to 3V3 PDN, $C_{19}$= $C_{20}$= $C_{30}$=10 uF, $C_{21}$=$C_{22}$=$C_{23}$=$C_{24}$=$C_{25}$=$C_{26}$= 0.1 uF.}
  \label{capacitors}
    \resizebox{1\textwidth}{!}{\begin{minipage}{\textwidth}
\begin{tabular}{|l|l|llllllll}
\cline{1-2}
 Exp. & Caps &                       &                       &                       &                       &                       &                       &                       &                       \\ \cline{1-3}
2 Caps & $c_{19}$ & \multicolumn{1}{l|}{$c_{20}$} &                       &                       &                       &                       &                       &                       &                       \\ \cline{1-4}
3 Caps & $c_{19}$ & \multicolumn{1}{l|}{$c_{20}$} & \multicolumn{1}{l|}{$c_{30}$} &                       &                       &                       &                       &                       &                       \\ \cline{1-6}
5 Caps & $c_{19}$ & \multicolumn{1}{l|}{$c_{20}$} & \multicolumn{1}{l|}{$c_{30}$} & \multicolumn{1}{l|}{$c_{21}$} & \multicolumn{1}{l|}{$c_{22}$} &                       &                       &                       &                       \\ \cline{1-8}
7 Caps &  $c_{19}$& \multicolumn{1}{l|}{$c_{20}$} & \multicolumn{1}{l|}{$c_{30}$} & \multicolumn{1}{l|}{$c_{21}$} & \multicolumn{1}{l|}{$c_{22}$} & \multicolumn{1}{l|}{$c_{23}$} & \multicolumn{1}{l|}{$c_{24}$} &                       &                       \\ \hline
9 Caps & $c_{19}$ & \multicolumn{1}{l|}{$c_{20}$} & \multicolumn{1}{l|}{$c_{30}$} & \multicolumn{1}{l|}{$c_{21}$} & \multicolumn{1}{l|}{$c_{22}$} & \multicolumn{1}{l|}{$c_{23}$} & \multicolumn{1}{l|}{$c_{24}$} & \multicolumn{1}{l|}{$c_{25}$} & \multicolumn{1}{l|}{$c_{26}$} \\ \hline
\end{tabular}
\end{minipage}}
\end{table}
\section{Results}\label{results}
\subsection{Emulating Tamper Events}\label{sec:emulation}
To emulate PCB tampering, various decoupling capacitors with different capacitances are added to the board.
Capacitors are chosen primarily due to three reasons. First, they play a major role in delivering power to the integrated circuits (ICs) on the PCB. Second, according to the ERAI~\cite{erai}, the capacitors are the most counterfeiting products in the market. Third, adding, removing or replacing any components on the PCB, e.g., implanting a spy chip, will cause changes in overall capacitance of the PDN, and therefore, such attacks type can be emulated by capacitors.

\begin{figure}[t!]
     	\centering \noindent
     	\includegraphics[width=0.8\linewidth]{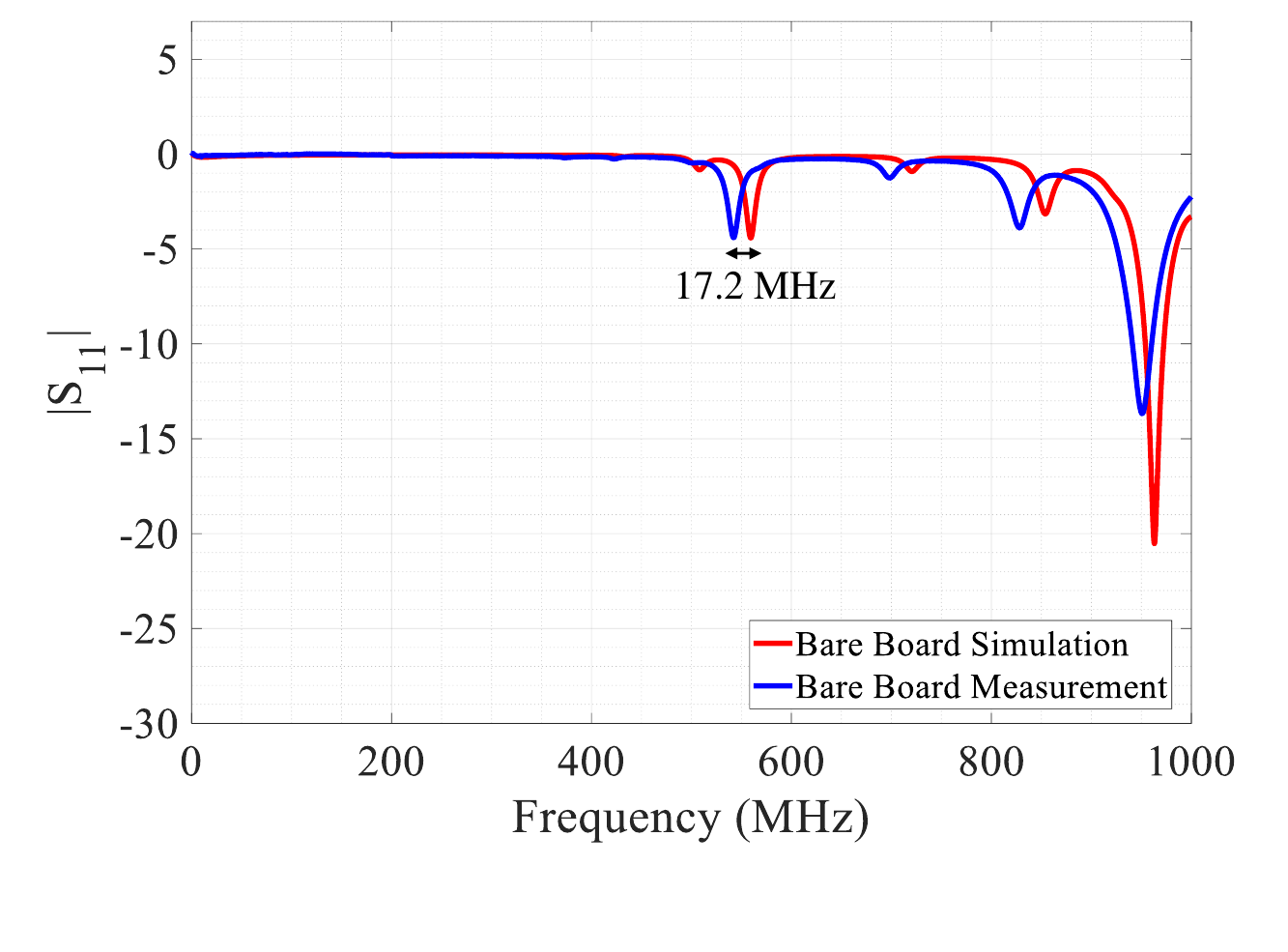}
      \vspace*{-5mm}
     	\caption{Simulated signature of $|S_{11}|$ extracted from ANSYS SIwave and measured $|S_{11}|$ of the bare board showing similar patterns with a shift in resonance frequencies.}
     	\label{Fig_9}
        \end{figure}

The process commenced with a bare board configuration, where no components were present.
The bare board layout was imported into the simulation software, enabling us to simulate the PCB and obtain the $|S_{11}|$ signature.
In Fig.~\ref{Fig_9}, we can observe the simulated and measured $|S_{11}|$ signature of the bare board, which exhibit good agreement.
However, there is a 17.2 MHz shift between them. 
This shift is solely due to the inevitable impurities in the substrate, as there are no components on the board.
Subsequently, at each stage, we added one or two capacitors to the PCB's PDN as shown in Table.~\ref{capacitors}.


\subsection{Identifying the Desired Frequency Band}
In our investigations, we acknowledge that different Trojans and tamper events exhibit detectable characteristics at different frequency bands, which vary depending on their size and area overhead. 
Specifically, smaller tampers, such as those at the IC level, tend to be detectable at very high frequencies (GHz bands)~\cite{mosavirik2023silicon}. 
However, our focus lies in the detection of attacks at the board level, indicating our interest in the MHz bandwidth.
The frequency bands of interest is the range in which the DTW analysis is applied to the results, allowing us to observe the impact of the components on the board. 
The resonance frequency of the circuit is highly sensitive to the addition or removal of components. 
Therefore, we chose the bandwidth such that the resonance frequency is at the center.
For high precision, we selected a range that spans 10\% around the lowest resonance frequency obtained from simulation, enabling us to specifically observe the effects of component addition or removal. 
For instance, when two capacitors are added to the circuit, the resonance frequency occurs at 234 MHz. 
Consequently, the DTW is applied to the frequency range of 222 MHz to 246 MHz. 
It is important to mention that for the bare board experiment, due to the absence of components on the board, we choose for a wide band comparison.
This implies that we compute the DTW distance of the second column from Table.~\ref{DTW distances-case1}, spanning a bandwidth of 1 MHz to 1 GHz.
To enhance the SNR and improve detection confidence while mitigating the impact of environmental noise and measurement uncertainties, we performed averaging on multiple measurements.

\subsection{Assessment of the Golden-free Tamper Detection Method}
Our methodology is validated through measurements performed on the board mentioned in Sect.~\ref{sec:dut}, where we employed a VNA channel probe connected to port J5 to accurately measure the $|S_{11}|$ of the PCB's 1.8V PDN.
Notably, no additional physical modifications were made to the system. 
Incorporating the approximate values of parasitic inductance and resistance in simulation tools enables more accurate predictions of the impedance behavior.
In most cases, obtaining the exact values is challenging.
Even the component vendors may not possess precise knowledge of these values.
To demonstrate the effectiveness of our approach, we performed extensive measurements on several proof-of-concept configurations.
The method is comprehensively discussed through two case studies. 

\begin{table}[!t]
     \centering
  \setlength{\tabcolsep}{0.5em}
 \renewcommand{\arraystretch}{0.07}

 \caption{Case Study 1: DTW distances using approximate values of ESR and ESL showing the impact of adding or removing components. Diagonal cell values represent lower DTW distances, illustrating golden detection with no components added or removed, (S) indicates simulation, (M) indicates measurement.} 
  \resizebox{0.7\textwidth}{!}{\begin{minipage}{\textwidth}
 \label{DTW distances-case1}
\begin{threeparttable}
\begin{tabular}{|c|c|c|c|c|c|c|c|c|c|}
 \hline
 \thead{$Reference \Rightarrow$ \\ $Test \Downarrow$}&\thead{Bare Board (S)} &\thead{2 Caps (S)} & \thead{ 3 Caps (S)} & \thead{ 5 Caps (S)} & \thead{ 7 Caps (S)} & \thead{ 9 Caps (S)}\\
  \hline
 \thead{Bare Board (M)}&\thead{\textcolor{deepgreen}{\textbf{189}}} &\thead{1247} & \thead{1250}& \thead{1316}& \thead{1295}& \thead{1264} \\
 \hline
 \thead{2 Caps (M)}&\thead{1291} &\thead{\textcolor{deepgreen}{\textbf{23.8}}} & \thead{33}& \thead{1316}& \thead{1270}& \thead{1239}\\  
 \hline
  \thead{3 Caps (M)}&\thead{1283} &\thead{41.8} & \thead{\textcolor{deepgreen}{\textbf{30}}}& \thead{1329}& \thead{1286}& \thead{1261}\\ 
  \hline
  \thead{5 Caps (M)}&\thead{1308} &\thead{1224} & \thead{1227}& \thead{\textcolor{deepgreen}{\textbf{21.3}}}& \thead{715}& \thead{1211}\\ 
  \hline
  \thead{7 Caps (M)}&\thead{1305} &\thead{1232} & \thead{1235}& \thead{60.2}& \thead{\textcolor{deepgreen}{\textbf{170}}}& \thead{1210}\\ 
  \hline
  \thead{9 Caps (M)}&\thead{1205} &\thead{1247} & \thead{1251}& \thead{1331}& \thead{1263}& \thead{\textcolor{deepgreen}{\textbf{212}}}\\ 
  \hline
 \end{tabular}
 \end{threeparttable}
 \end{minipage}}
\end{table}

\subsubsection{Case Study 1: Addition or Removal of Components}
In this case study, we assess the feasibility of detecting addition or removal of a component from the PCB.
To emulate the tampering, we added multiple capacitors in various trials to the board.
To generate the simulated golden signature for each trial, the verifier sets the expected values for parasitic impedance of the PCB's component.
As shown in Table.~\ref{DTW distances-case1}, if we consider the first row as the reference signature obtained from the simulation data and the first column as the measured signatures corresponding to each of the experiments, the values in the diagonal cells represent the DTW distances between the simulation and measurement data of each experiment.
As it can be observed the simulated and measured $|S_{11}|$ signatures of identical configurations exhibit lower DTW distances compared to cases, where the simulated configuration differs from the physical sample configuration.


\begin{table}[h]
\centering
\caption{Case Study 2: DTW distances using deviating values of ESL and ESR to emulate the replacement genuine parts with counterfeit ones showing significantly larger DTW distances than the diagonal cell values in the Table.~\ref{DTW distances-case1}.}
\label{DTW distances-case2}
\begin{tabular}{|c|c|c|c|c|c|}
\hline
\thead{Sim vs Meas} & \thead{2 Caps} & \thead{3 Caps} & \thead{5 Caps} & \thead{7 Caps} & \thead{9 Caps} \\
\hline
\thead{DTW Distances} & \thead{\textcolor{red}{{1100}}} & \thead{\textcolor{red}{{1095}}} & \thead{\textcolor{red}{{1205}}} & \thead{\textcolor{red}{{1127}}} & \thead{\textcolor{red}{{856}}} \\
\hline
\end{tabular}
\end{table}
\subsubsection{Case Study 2: Replacing Parts with Counterfeit Ones}
In this case study, we assess the feasibility of detecting a component that has been replaced with a counterfeit part.
The assumption here is that the counterfeit part has a significant parasitic impedance deviation.
Since we did not have access to counterfeit components, we edited instead the ESL and ESR values in our simulations.
We chose deviations in order of 10 and 1.3 for our ESL and ESR values.
Such factors were obtained by averaging the existing ESL and ESR values of similar components from various vendors mentioned in their datasheets.
As it can be observed in Table.~\ref{DTW distances-case2}, the measured and simulated signatures exhibit substantial disparities to the point where the DTW distance between them becomes very large compared to the distances of the previous case study, where no components were added or removed.
This confirms that replacing components with fake parts could be detected if the fake parts have a different parasitic behavior.

\subsubsection{Accuracy of the Proposed Method}
The case studies offers valuable insights into the method's sensitivity to the parasitic values.
To gain a deeper insight into the method, the verifier can introduce a margin factor, denoted as $\eta$. 
This factor illustrates the difference in the DTW distance value between an untampered board and the smallest DTW distance observed in all experiments. 
Ideally, this margin should be zero.
However, manufacturing process variations lead to a non-zero value, and poorly selected ESL and ESR values further increase $\eta$. 
The smaller the margin factor, the better our method becomes at detecting minor tamper events, enhancing its effectiveness. 
Consequently, the sufficiency of the ESL and ESR choice determines the types of tamper events we aim to detect.
Detecting more sophisticated tampering requires greater accuracy in the ESL and ESR values.

\subsection{Validation of the Tamper Detection Across Different PDNs}
In this section, we discuss the robustness of the proposed method in detecting tampering events connected to other adjacent PDNs. 
When electric currents are directed through one PDN, magnetic fields are generated, which have the potential to induce voltages within neighboring PDNs due to the principle of mutual inductance.
The mutual coupling between PDNs allows us to detect alterations in the impedance within one PDN by leveraging another PDN. 
This becomes especially valuable in situations where access is typically limited to a single PDN, yet we maintain the ability to identify tampers associated with other PDNs.
In our experimental investigation, the PDN under test is the 1V8 PDN.
In the third experiment in Sect.\ref{sec:emulation}, we added the $C_{30}$ connected to the 3V3 PDN to the board.
The introduction of $C_{30}$ to the circuit brought about changes in the $|S_{11}|$ parameter, consequently affecting the DTW distances as well.
The obtained results are presented in Table.~\ref{DTW distances-case1}, where it can be observed that the DTW distances between the second and third experiments differ.
Although the observed difference is not as pronounced as the values associated with components directly connected to the PDN under test, the changes in the DTW distances remain detectable.

\section{Conclusion}
In this paper, we explored the feasibility of using simulated golden signatures as a substitute for the physical ones.
Our results validate the efficacy of the proposed detection method, which relies on configuring the expected parasitic values during simulation.
Given approximated values for the parasitic impedance of components on the PCB, replacing the physical golden sample with its simulated signature is feasible.
Moreover, the proposed approach not only detects potential attacks on specific PDNs of the PCB but also extends its capabilities to identify attacks on other PDNs, proving particularly valuable when access to all PDNs is unavailable.
We believe this work lays the foundation for improving the proposed method for other real-world scenarios, such as the golden-free verification of much larger systems with hundreds of components, detection of combined tampering and counterfeiting attacks using more advanced machine learning algorithms, and defining reliable thresholds based on the parasitic values.

\section*{Acknowledgment}
This work was sponsored by Electric Power Research Institute (EPRI).




\bibliographystyle{ieeetr}
\bibliography{references}

\begin{thebibliography}{10}

\bibitem{big_hack}
J.~Robertson and M.~Riley, ``{The Big Hack: How China used a Tiny Chip to Infiltrate US Companies},'' {\em Bloomberg Businessweek}, vol.~4, 2018.

\bibitem{fake_cisco}
D.~Janushkevich, ``{The Fake Cisco: Hunting for Backdoors in Counterfeit Cisco Devices},'' {\em F-Secure Consulting, Hardware Security Team}, 2020.

\bibitem{botero2021hardware}
U.~J. Botero, R.~Wilson, H.~Lu, M.~T. Rahman, M.~A. Mallaiyan, F.~Ganji, N.~Asadizanjani, M.~M. Tehranipoor, D.~L. Woodard, and D.~Forte, ``Hardware trust and assurance through reverse engineering: A tutorial and outlook from image analysis and machine learning perspectives,'' {\em ACM Journal on Emerging Technologies in Computing Systems (JETC)}, vol.~17, no.~4, pp.~1--53, 2021.

\bibitem{chaudhary2017automatic}
V.~Chaudhary, I.~R. Dave, and K.~P. Upla, ``Automatic visual inspection of printed circuit board for defect detection and classification,'' in {\em 2017 International Conference on Wireless Communications, Signal Processing and Networking (WiSPNET)}, pp.~732--737, IEEE, 2017.

\bibitem{piliposyan2022pcb}
G.~Piliposyan and S.~Khursheed, ``{PCB} hardware {T}rojan run-time detection through machine learning,'' {\em IEEE Transactions on Computers}, 2022.

\bibitem{mosavirik2022scatterverif}
T.~Mosavirik, F.~Ganji, P.~Schaumont, and S.~Tajik, ``Scatterverif: Verification of electronic boards using reflection response of power distribution network,'' {\em ACM Journal on Emerging Technologies in Computing Systems (JETC)}, vol.~18, no.~4, pp.~1--24, 2022.

\bibitem{werner2022detection}
F.~T. Werner, M.~Prvulovic, and A.~Zaji{\'c}, ``Detection of recycled {ICs} using backscattering side-channel analysis,'' {\em IEEE Transactions on Very Large Scale Integration (VLSI) Systems}, vol.~30, no.~9, pp.~1244--1255, 2022.

\bibitem{zhu2023pdnpulse}
H.~Zhu, H.~Shan, D.~Sullivan, X.~Guo, Y.~Jin, and X.~Zhang, ``{PDNP}ulse: sensing {PCB} anomaly with the intrinsic power delivery network,'' {\em IEEE Transactions on Information Forensics and Security}, 2023.

\bibitem{safa2023counterfeit}
M.~S. Safa, T.~Mosavirik, and S.~Tajik, ``Counterfeit chip detection using scattering parameter analysis,'' in {\em 2023 26th International Symposium on Design and Diagnostics of Electronic Circuits and Systems (DDECS)}, IEEE, 2023.

\bibitem{mosavirik2023impedanceverif}
T.~Mosavirik, P.~Schaumont, and S.~Tajik, ``Impedanceverif: On-chip impedance sensing for system-level tampering detection,'' {\em IACR Transactions on Cryptographic Hardware and Embedded Systems}, pp.~301--325, 2023.

\bibitem{bhattacharyay2022vipr}
A.~Bhattacharyay, P.~Chakraborty, J.~Cruz, and S.~Bhunia, ``{VIPR-PCB}: a machine learning based golden-free {PCB} assurance framework,'' in {\em Proceedings of the 59th ACM/IEEE Design Automation Conference}, pp.~793--798, 2022.

\bibitem{chowdhury2023golden}
A.~B. Chowdhury, A.~Mahapatra, Y.~Liu, P.~Krishnamurthy, F.~Khorrami, and R.~Karri, ``A golden-free approach to detect {T}rojans in {COTS} multi-{PCB} systems,'' {\em IEEE Micro}, 2023.

\bibitem{krishnamurthy2022multi}
P.~Krishnamurthy, V.~R. Surabhi, H.~Pearce, R.~Karri, and F.~Khorrami, ``Multi-modal side channel data driven golden-free detection of software and firmware {T}rojans,'' {\em IEEE Transactions on Dependable and Secure Computing}, 2022.

\bibitem{vintsyuk1968speech}
T.~K. Vintsyuk, ``Speech discrimination by dynamic programming,'' {\em Cybernetics}, vol.~4, no.~1, pp.~52--57, 1968.

\bibitem{sakoe1978dynamic}
H.~Sakoe and S.~Chiba, ``Dynamic programming algorithm optimization for spoken word recognition,'' {\em IEEE transactions on acoustics, speech, and signal processing}, vol.~26, no.~1, pp.~43--49, 1978.

\bibitem{ansys_siwave_2023}
{ANSYS, Inc.}, ``{ANSYS SIwave 2023 R2}.'' \url{http://www.ansys.com}, 2023.

\bibitem{siwave}
B.~Wei and S.~G.~P. Jr, ``New integrated workflow for {EMI} simulation,'' {\em APEMC 2015}.

\bibitem{erai}
{Akhoundov, Damir}, ``{2019 ERAI Reported Parts Statistics},'' {\em ERAI Blog}, 2020.

\bibitem{mosavirik2023silicon}
T.~Mosavirik, S.~K. Monfared, M.~S. Safa, and S.~Tajik, ``Silicon echoes: Non-invasive {T}rojan and tamper detection using frequency-selective impedance analysis,'' {\em IACR Transactions on Cryptographic Hardware and Embedded Systems}, pp.~238--261, 2023.

\end{thebibliography}

\end{document}